\documentclass[sigconf]{acmart}
\usepackage{soul}
\usepackage{multirow}
\usepackage{enumitem}
\usepackage[ruled,linesnumbered]{algorithm2e}
\usepackage{marvosym}
\usepackage{subfigure}
\AtBeginDocument{%
 }

\setcopyright{acmcopyright}
\copyrightyear{2023}
\acmYear{2023}
\setcopyright{acmlicensed}\acmConference[CIKM '23]{Proceedings of the 32nd ACM International Conference on Information and Knowledge Management}{October 21--25, 2023}{Birmingham, United Kingdom}
\acmBooktitle{Proceedings of the 32nd ACM International Conference on Information and Knowledge Management (CIKM '23), October 21--25, 2023, Birmingham, United Kingdom}
\acmPrice{15.00}
\acmDOI{10.1145/3583780.3615137}
\acmISBN{979-8-4007-0124-5/23/10}
\usepackage{color, xspace}

\begin{document}
\title{HAMUR: Hyper Adapter for Multi-Domain Recommendation}

\author{Xiaopeng Li}
\affiliation{%
 \institution{City University of Hong Kong}
 \city{Hong Kong}
 \country{China}
}
\email{xiaopli2-c@my.cityu.edu.hk}

\author{Fan Yan}
\affiliation{%
 \institution{Huawei Noah’s Ark Lab}
 \city{Shenzhen}
 \country{China}
}
\email{yanfan6@huawei.com}

\author{Xiangyu Zhao\textsuperscript{\Letter}}
\affiliation{%
 \institution{City University of Hong Kong}
 \city{Hong Kong}
 \country{China}
 }
\thanks{{\Letter} Corresponding authors.}
\email{xianzhao@cityu.edu.hk}

\author{Yichao Wang}
\affiliation{%
 \institution{Huawei Noah’s Ark Lab}
 \city{Shenzhen}
 \country{China}
}
\email{wangyichao5@huawei.com}

\author{Bo Chen}
\affiliation{%
 \institution{Huawei Noah’s Ark Lab}
 \city{Shenzhen}
 \country{China}
}
\email{chenbo116@huawei.com}

\author{Huifeng Guo\textsuperscript{\Letter}}
\affiliation{%
 \institution{Huawei Noah’s Ark Lab}
 \city{Shenzhen}
 \country{China}
}
\email{huifeng.guo@huawei.com}

\author{Ruiming Tang}
\affiliation{%
 \institution{Huawei Noah’s Ark Lab}
 \city{Shenzhen}
 \country{China}
}
\email{tangruiming@huawei.com}
\renewcommand{\shortauthors}{Xiaopeng Li et al.}

\begin{abstract} 
Multi-Domain Recommendation (MDR) has gained significant attention in recent years, which leverages data from multiple domains to enhance their performance concurrently.
However, current MDR models are confronted with two limitations.
Firstly, the majority of these models adopt an approach that explicitly shares parameters between domains, leading to mutual interference among them.
Secondly, due to the distribution differences among domains, the utilization of static parameters in existing methods limits their flexibility to adapt to diverse domains. To address these challenges, we propose a novel model \textbf{H}yper \textbf{A}dapter for \textbf{MU}lti-Domain \textbf{R}ecommendation (\textbf{HAMUR}). 
Specifically, HAMUR consists of two components: (1). Domain-specific adapter, designed as a pluggable module that can be seamlessly integrated into various existing multi-domain backbone models, and (2). Domain-shared hyper-network, which implicitly captures shared information among domains and dynamically generates the parameters for the adapter.
We conduct extensive experiments on two public datasets using various backbone networks. The experimental results validate the effectiveness and scalability of the proposed model. And we release our code implementation publicly \footnote{https://github.com/Applied-Machine-Learning-Lab/HAMUR}\footnote{https://gitee.com/mindspore/models/tree/master/research/recommend/HAMUR}.
\end{abstract}


\maketitle

\section{Introduction}

Click-through rate~(CTR) prediction is crucial for recommender systems, which find extensive application in e-commerce, music radio, social media, and online recommendations, among others~\cite{davidson2010youtube, okura2017embedding}. Traditional CTR models typically concentrate on a single domain~\cite{zhao2018deep,zhao2018recommendations,covington2016deep,li2023automlp}, limiting data collection and model training in that specific domain, and it is also faced with two limitations, data sparseness, and cold start problems. Cross-Domain Recommendation (CDR) approaches have been proposed, which make use of the extensive data available in the source domain to supplement the target domain. However, with the business growth and the increasing demand for personalized recommendations, models must cater to multiple diverse domains within the system. But due to the varying distribution characteristics of different domains, simply merging multi-domain data would introduce domain bias issues~\cite{chang2023pepnet}. On the other hand, constructing separate models for each domain would impose significant computational and manual maintenance costs~\cite{wang2022causalint,wang2023plate}. Therefore, the critical research challenge lies in designing a unified framework that efficiently integrates multiple domains while optimizing overall predictive performance. 

In fact, the main challenge in multi-domain recommendation is to strike a balance between inter-domain sharing and intra-domain independence~\cite{sheng2021one,gao2023autotransfer}. Existing multi-domain recommendation models fall into two categories: (1). Hard sharing models, which typically employ the multi-task learning paradigm, sharing information in the bottom layers while designing individual domain towers at the top~\cite{shen2021sar,ma2018your}. Other hard-sharing strategies, such as STAR~\cite{sheng2021one}, utilize a star topology, where one network is shared among different domains and combined with different domain-specific networks through element-wise multiplication. (2). Soft-sharing models, which indirectly share domain parameters among domains by incorporating auxiliary modules, such as cells, units, or layers, into the main network to carry domain-specific information~\cite{zhang2022leaving, jiang2022adaptive}. For example, domain information is shared through gating networks in the PEPNet model~\cite{chang2023pepnet}, where a gate neural unit is proposed to amplify valid signals, while M2M~\cite{zhang2022leaving} introduces meta units generated based on specific scenario knowledge. 
On the other hand, some models adopt various fusion strategies, e.g., ADI~\cite{jiang2022adaptive} employs a fusion layer that concatenates domain-related information generated by both shared and specific networks.
Despite the effectiveness demonstrated by the aforementioned approaches in MDR, they still encounter two limitations. 
Firstly, relying on explicit sharing, particularly hard-sharing, can lead to significant bias in certain domains, as the parameter-sharing paradigm is artificially prioritized. Secondly, both hard-sharing and soft-sharing methods employ static parameters, which exhibit limited scalability and cannot ensure customized modeling patterns for diverse domains. 
Consequently, it is imperative to explore alternative approaches that incorporate dynamic sharing parameters so as to enable flexible adaptation to rapidly evolving domains and accommodating varying distributions among different domains.
\begin{figure}[t]
  \centering
  \includegraphics[width = 0.9\linewidth]{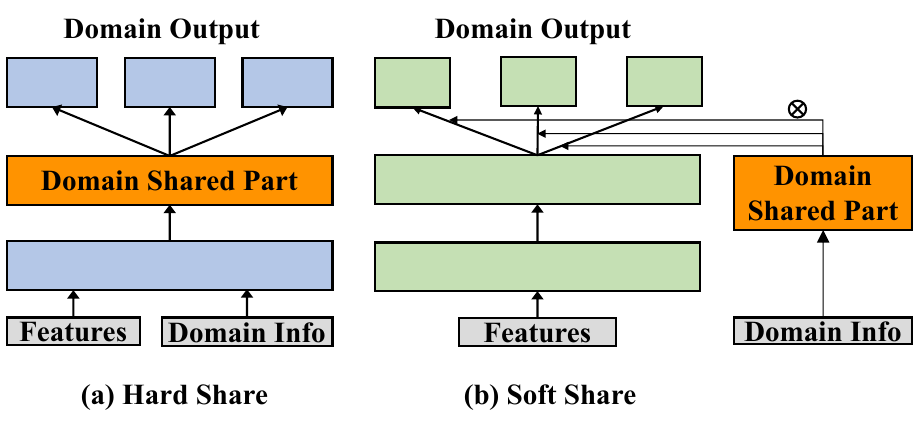}
  \vspace{-3mm}
  \caption{Existing Multi-domain sharing methods. }
  \Description{Existing Multi-domain sharing methods.}
  \label{fig:ShareMethod}
\end{figure}

In this paper, we propose Hyper Adapter for MUlti-domain Recommendation (HAMUR) to address the above challenges of multi-domain adaptation through adapter technique in the natural language processing (NLP) field~\cite{houlsby2019parameter}. However, considering the disparities in data characteristics between NLP and recommendations, the model's parameters necessitate frequent updates in the recommendations system, which could be labor-intensive if adjusting the adapter for each fine-tuning iteration. Therefore, we refrain from using the two-stage pre-training and fine-tuning approach. Instead, we directly integrate the adapter into the backbone model, employing end-to-end training to facilitate dynamic adaptation to changing data distributions during online deployment. Additionally, we design a hyper-network that is shared among different domains, implicitly capturing domain-specific patterns and dynamically generating the weights of adapters. The input of the hyper-network is the instance-level embedding that carries domain-specific information. We validate the effectiveness of our proposed HAMUR against state-of-art MDR models on two public datasets. 
The main contributions of this work can be summarized as follows:
\begin{itemize}[leftmargin=*]
\item We propose HAMUR, a novel solution to tackle the challenge of predicting click-through rates (CTR) in a multi-domain setting. Our approach seamlessly integrates with various existing backbones as a plug-and-play component;
\item To account for the variations in data distribution across different domains, we propose a hyper-network that is shared among all domains, generating adapter weight parameters dynamically. This strategy enables the adapter to quickly capture domain-specific patterns, leading to robust generalization across all domains. Furthermore, the hyper-network can implicitly capture shared information across domains;
\item Based on experimental results on two public datasets, our approach surpasses existing state-of-the-art methods in terms of effectiveness. Furthermore, we evaluate the effectiveness of our method by substituting diverse backbone networks, demonstrating the scalability of HAMUR.
  \end{itemize}

\section{Methodology}

In this section, we begin by formulating the task of multi-domain click-through rate (CTR) prediction. We then provide an overview of the architecture of our proposed method. Subsequently, we present a detailed description of our method, which includes the domain adapter cell and the hyper-network. In the end, we will discuss task optimization to conclude this section. 

\subsection{Problem Formation \label{sec:Problem Formation}}
 This paper focuses on the CTR prediction task in Multi-Domain Recommendation (MDR). Traditional CTR prediction models are trained on a single domain, where various distinguishing features such as user profiles, historical behaviors, item features, and context features are taken as input $\boldsymbol{x}$. After passing through the embedding layer, the raw features are then mapped into low-dimensional embedding vectors $\boldsymbol{e}$, which will be fed into the prediction model afterward, and the predicted CTR will be obtained. The label $y \in \{0, 1\}$ indicates whether a click is occurred or not.
In contrast to single-domain prediction, an MDR task involves a new feature $\boldsymbol{p}\in \{1,..., D\}$, where $\boldsymbol{p}$ represents the domain indicator for samples from $D$ domains. The objective of CTR in MDR is to construct a unified model $\mathcal{F}$ that can accurately predict  $\hat{p}$ for $D$ domains simultaneously. It could be written as:
\begin{equation}
    \hat{p}  = \mathcal{F} (\boldsymbol{x}, \boldsymbol{p};\theta)
\end{equation}
for all $D$ domains, the problem can be formulated as:
\begin{equation}
    \min_\theta \frac{1}{|D|} \sum_d\sum_{i=1}^{|D_d|} \mathcal{L}_{CTR}(y^i,\hat{p}^i_d)
\end{equation}
where $d$ refers to the $d$-th domain, $\theta$ refers to the parameters of MDR model $\mathcal{F}$, and our objective is to minimize the loss function $\mathcal{L}_{CTR}$ to get the best MDR model $\mathcal{F}$ for the recommender system.

\subsection{Framework Overview}
\begin{figure*}[ht]
  \centering
 \includegraphics[width = 0.8\linewidth]{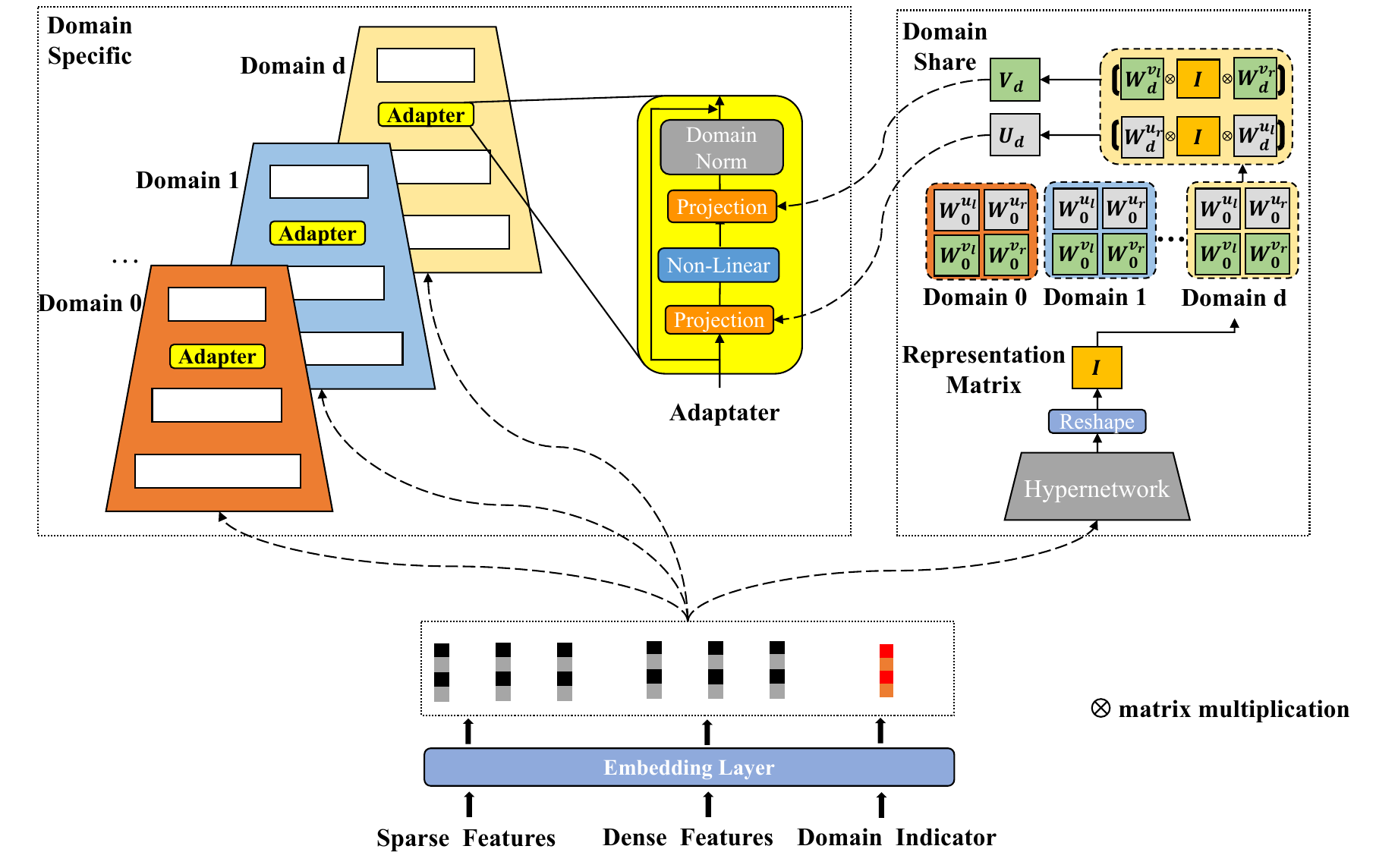}
 \caption{An illustration of the overall architecture of our proposed model. 
 }
 \label{fig:Framework}
\end{figure*}
The structure of our model is shown in Figure~\ref{fig:Framework}. Our approach integrates the adapter and dynamic weight generation techniques together for MDR modeling. Specifically, we propose a domain-specific adapter cell that can be seamlessly incorporated into existing MDR models. Additionally, to capture the underlying similarities among different domains, we propose a domain-shared hyper-network to generate parameters dynamically for the adapter.

\begin{itemize}[leftmargin=*]
\item \textbf{Domain-specific adapter cell}: To capture the domain personality, we propose to utilize a small network named adapter as a pluggable component for MDR modeling. It is designed as a bottleneck shape component that comprises four layers, a down projection layer,  a non-linear layer, an up projection layer, and a domain normalization layer, besides a skip connecting established from the beginning to the end of the adapter. Compared with the adapter~\cite{houlsby2019parameter} in the field of NLP, which is used for fine-tuning Large pre-trained Language Models~(LLM). We have forsaken the two-stage approach and instead integrated the adapter directly into the backbone network to facilitate end-to-end training. We use the adapter as a pluggable component, which could be easily integrated into various backbone networks such as MLP~\cite{amari1967theory}, DCN~\cite{wang2017deep}, Wide \& Deep~\cite{cheng2016wide}, DeepFM~\cite{guo2017deepfm}, and so on.

\item \textbf{Domain-shared hyper-network}: In order to address the issue of domain bias, we present a solution in the form of a shared hyper-network, which is used for generating parameters for the domain adapters. To improve computational efficiency, the matrix low-rank decomposition method is employed, whereby the target matrix is decomposed into the product of three matrices. Specifically, the hyper-network input comprises instances with domain information $(\boldsymbol{x},\boldsymbol{p})$ and generates an instance-leveled domain representation matrix $\boldsymbol{I}$. Matrices $\boldsymbol{W}^{u_l}_d$, $\boldsymbol{W}^{u_r}_d$, $\boldsymbol{W}^{v_l}_d$, and $\boldsymbol{W}^{v_r}_d$ are then multiplied with the matrix $\boldsymbol{I}$ to obtain matrices $\boldsymbol{U}_d$ and $\boldsymbol{V}_d$, which serve as adapter's parameters for domain $d$.
\end{itemize}
\vspace{-3mm}
\subsection{Domain-Specific Adapter Cell}

The domain-specific adapter cell has been designed to capture each domain's unique characteristics. In this section, the structure of domain-specific adapter cells and domain adapter normalization layer will be introduced subsequently.

\subsubsection{Domain Adapter Layers}

In the field of NLP, fine-tuning involves the insertion of small modules called adapters into pre-trained models without any modifications to the architecture or parameters of the original model. However, frequently updating model parameters is necessary in the context of recommender systems. This is particularly evident in scenarios such as e-commerce websites where promotional activities are carried out or in video websites where new videos are released. In order to align with these changes, the model parameters must be updated accordingly.
If we persist with the two-stage adjustment approach for the adapter in each model update, it will be time-consuming and labor-intensive, which is unacceptable.
So we design the adapter as a plug-in into the existing multi-domain model directly, thereby enabling its participation in end-to-end training.

The structure of the adapter is shown in Figure~\ref{fig:Framework}. The adapter cell, denoted as $\boldsymbol{A}_d$, is designed with the shape of the bottleneck. It includes four layers, a down-projection $\boldsymbol{U}_d \in \mathbb{R}^{h \times s}$, a sigmoid non-linear layer $\sigma$, an up-projection $\boldsymbol{V}_d \in \mathbb{R}^{s \times h}$, and a domain normalization layer, besides a residual connection from the beginning of the cell to the end. $h$ denotes the input dimension, and $s$ denotes the bottleneck dimension, $s<<h$. Through down-projection, input data is compressed into low-dimensional embedding. It is followed by a non-linear layer and feature reconstruction via up-projection.
The mathematical formulation for the whole domain-specific adapter cell is defined as:
\begin{equation}
\label{eq:eq1}
    \boldsymbol{A}_d (\boldsymbol{x}) = DN_d \Big ( \boldsymbol{V}_d (\sigma (\boldsymbol{U}_d (\boldsymbol{x}))) \Big) +\boldsymbol{x}
\end{equation}
where $\boldsymbol{x}$ represents the hidden input state, $DN_d$ denotes domain normalization and will be elaborated upon in the subsequent section. $\boldsymbol{U}_d$ and $\boldsymbol{V}_d$ are generated through a shared hyper-network, which will be described in Section~\ref{sec:Domain Shared Hyper Network}, and $d$ represents the $d$-th domain.

\subsubsection{Domain Adapter Normalization}
The normalization layer in the adapter design conventionally utilizes layer normalization. In the NLP field, the length of the input sequence varies, and there is a semantic correlation between individual tokens in one sentence. Hence, it is reasonable to use layer normalization to normalize the tokens in each sentence. However, this approach is not applicable in the area of recommender system, where the embeddings of distinct feature fields represent different meanings. Directly normalizing across feature fields would be an irrational practice. Consequently, we opt for batch normalization to normalize the instances of the entire mini-batch.

In the case of multi-domain CTR prediction, different from single-domain tasks, the data distribution is locally Independent Identically Distributed (IID) in each individual domain. Hence, we propose the use of domain normalization, which is expressed as:
\begin{equation}
    DN_d = \gamma_d \odot \frac{ \boldsymbol{x} - \mu}{\sqrt{\sigma^2+ \epsilon}} + \beta_d
\end{equation}
where \ $\gamma_d$ and $\beta_d$ are learnable parameters for domain-specific scale and bias parameters. $\odot$ is the element-wise multiplication. $\mu$ and $\sigma$ represent the mean and standard deviation of the data $\boldsymbol{x}$.

\subsection{Domain Shared Hyper-Network \label{sec:Domain Shared Hyper Network}}

In order to capture implicit information shared between different domains and dynamically generate parameters, we propose the inter-domain sharing hyper-network. Additionally, to improve the generation efficiency of the target matrix, we draw inspiration from~\cite{yan2022apg} and adopt a low-rank decomposition method to decompose the target matrix into the product of three matrices. During the generation process, the hyper-network generates only a core expression matrix $\boldsymbol{I}$, which reduces memory usage and significantly improves operational efficiency. The following section will provide a detailed description.

\subsubsection{Parameters Generation\label{subsec:Parameters Generation}}

This subsection will elaborate on how we use hyper-network to generate parameters dynamically. The input includes two parts, the raw features $\boldsymbol{x}$, and the domain indicator $\boldsymbol{p}$. They will be fed into the embedding layer. A mathematical expression could be formulated as:
\begin{equation}
\label{eq:eq3}
    \boldsymbol{Z} = emb \big ( \boldsymbol{x},\boldsymbol{p} \big)
\end{equation}
It is worth noting that our generation process is in instance-level. Therefore, for the convenience of description, we rewrite Equation~\eqref{eq:eq3}. Assuming our current batch size is ${B}$, then $\boldsymbol{Z}$ could be rewritten as:
\begin{equation*}
 \begin{aligned}
  \boldsymbol{Z} &= [\boldsymbol{z}^1,...,\boldsymbol{z}^i,..., \boldsymbol{z}^B ]\\
   &= [emb ({\boldsymbol{x}^1},\boldsymbol{p}^1),...,emb ({\boldsymbol{x}^i},\boldsymbol{p}^i),...,emb ({\boldsymbol{x}^B},\boldsymbol{p}^B) ]
 \end{aligned}
\end{equation*}
The domain-shared hyper-network $\mathcal{H}(.)$ is a shallow neural network. For each instance, $\boldsymbol{z}^i$, it will be fed into the domain-shared hyper-network $\mathcal{H}(.)$. After that, a one-dimensional representation $\boldsymbol{h}^i$ will be obtained, an expression containing different domain commonality. Then, we reshape $\boldsymbol{h}^i$ to form the matrix $\boldsymbol{I}^i$ to facilitate matrix operations in the subsequent stages. formally, this process is expressed as:
\begin{equation}
\label{eq:eq4}
  \boldsymbol{h}^i = \mathcal{H} (\boldsymbol{z}^i) 
\end{equation}
\begin{equation}
\label{eq:eq5}
  \boldsymbol{I}^i = reshape (\boldsymbol{h}^i)
\end{equation}
For  $\boldsymbol{I}^i$, we name it the representation matrix for the $i$-th instance.

\subsubsection{Low-Rank Decomposition}
This part will explain how we leverage low-rank matrix decomposition method to enhance computational efficiency. As previously mentioned, we generate one matrix $\boldsymbol{I}$ for each instance, but we require two target matrices, $\boldsymbol{U}_d$ and $\boldsymbol{V}_d$, so we utilize the low-rank decomposition method to address this issue, drawing inspiration from other literatures~\cite{yan2022apg}. Specifically, for the target matrices $\boldsymbol{U}_d^i \in \mathbb{R}^{s \times h}$ and $\boldsymbol{V}_d^i \in \mathbb{R}^{h \times s}$, we decompose these two matrices into three matrices multiplication, respectively, expressed as:
\begin{equation}
\label{eq:eq6}
    \boldsymbol{U}^i_d = \boldsymbol{W}_d ^{u_l} \cdot \boldsymbol{I}^i \cdot \boldsymbol{W}_d ^{u_r}
\end{equation}
\begin{equation}
\label{eq:eq7}
\boldsymbol{V}^i_d = \boldsymbol{W}_d ^{v_l} \cdot \boldsymbol{I}^i \cdot \boldsymbol{W}_d ^{v_r}
\end{equation}
where $\boldsymbol{W}_d ^{u_l} \in \mathbb{R}^ {s\times k}$, $\boldsymbol{W}_d ^{u_r} \in \mathbb{R}^ {k\times h}$, $\boldsymbol{W}_d ^{v_l} \in \mathbb{R}^ {h\times k}$, $\boldsymbol{W}_d ^{v_r} \in \mathbb{R}^ {k\times s}$, and $\boldsymbol{I}^i \in \mathbb{R}^{k \times k}$. During model training, $\boldsymbol{W}_d ^{u_l} $, $\boldsymbol{W}_d ^{u_r} $, $\boldsymbol{W}_d ^{v_l} $, and $\boldsymbol{W}_d ^{v_r} $ are the inherent parameters of the model that participate in the training process, thus we only need to generate the matrix $\boldsymbol{I}^i$ via the hyper-network. 

There are two primary reasons for utilizing low-rank decomposition. Firstly, it improves computational efficiency. By generating only one matrix $\boldsymbol{I}^i$ with a size of $k \times k$ directly, as opposed to directly generating the target matrices $\boldsymbol{U}^i_d$ and $\boldsymbol{V}^i_d$, we can significantly improve the operating efficiency. Secondly, the design enables the parameters $\boldsymbol{W}_d ^{u_l} $, $\boldsymbol{W}_d ^{u_r} $, $\boldsymbol{W}_d ^{v_l} $, and $\boldsymbol{W}_d ^{v_r} $ in the $d$-th domain to share domain-specific information across instances, thereby effectively capturing private domain patterns.

\subsection{Integration with Backbone Model}

This section will introduce how the adapter integrates into the backbone model.
As we mentioned in Section~\ref{sec:Problem Formation}, the base MDR model is denoted as:
\begin{equation*}
    \mathcal{F}(\theta) = [\mathcal{F}_1(\theta_1),...,\mathcal{F}_d(\theta_d),...,\mathcal{F}_D(\theta_D) ]
\end{equation*}
where $\mathcal{F}_d(\theta_d)$ denotes $d$-th domain model, and it could be any kind of existing backbone network, like MLP~\cite{amari1967theory}, DCN~\cite{wang2017deep}, Wide \& Deep~\cite{cheng2016wide}, and for all the $D$ models, they are domain independent. 

Upon integration with the domain-specific adapter cell, the model can be represented as $\mathcal{F}_d(\theta_d, \boldsymbol{A}_d)$, where the adapter is incorporated as an integral part of the model. It is important to note that these $D$ models can no longer be regarded as domain-independent since the adapters are generated through a domain-shared hyper-network. Technically, we chose to insert the adapter at the top level of the backbone network. It was motivated by the hierarchical nature of the model architecture. The lower layers of the model are responsible for feature extraction and intersection, while the upper layers produce a more refined feature mapping and the final prediction. Given our objective of facilitating rapid adaptation of the adapter to new domains, it would be counterproductive to introduce changes to the feature distribution prematurely, as this would weaken the feature extraction capability. Therefore, placing the adapter at the bottom of the model would be inappropriate. Overall, for the $i$-th instance, the output of CTR could be formulated as:
\begin{equation}
\label{eq:eq10}
    \hat{y}^i_d =\mathcal{F}_d(\boldsymbol{x}^i;\theta_d,\boldsymbol{A}^i_d) 
\end{equation} 

\subsection{Optimization}

The overall optimization goal of our model is:
\begin{equation}
\label{eq:eq8}
\mathcal{L}_{CTR} =   - {y}_i^d \log (\hat{y}_i^d ) - (1-{y}_i^d)\log (1-\hat{y}_i^d)
\end{equation}
The objective function applied in our model is the cross entropy loss function. $\hat{y}_i^d$ and ${y}_i^d$ denotes the prediction and the ground truth in the $d$-th domain, respectively. 

\begin{algorithm}[t]
\caption{\label{alg:HAMUR} The Algorithm for HAMUR.}\
\KwIn{features $\boldsymbol{x}$,  domain indicator $\boldsymbol{p}$, ground-truth labels $\boldsymbol{y}$, batch size $B$.}
\KwOut{Prediciton click label $\hat{y}$.}
 \While{not converged}{
  Sample a mini-batch data $(\boldsymbol{x},\boldsymbol{p},y)$ with size $B$\;
  Get $\boldsymbol{Z}$ via Equation~\eqref{eq:eq3}\;
  \ForEach{$\boldsymbol{z}^i \in \boldsymbol{Z}$}{

    \ForEach{$d$ = $1$ to $D$}{
    Get domain parameters $\boldsymbol{W}_d^{u_l}$, $\boldsymbol{W}_d ^{u_r}$, $\boldsymbol{W}_d ^{v_l} $, $\boldsymbol{W}_d^{v_r} $\;
    Get domain-specific backbone  $\mathcal{F}_d(\theta_d)$\;
         \eIf{$p^i$==$d$}{
         Calculate  $\boldsymbol{I^i}$ via Equation~\eqref{eq:eq4},~\eqref{eq:eq5}\;
         Calculate $\boldsymbol{U}^i_d$ and $\boldsymbol{V}^i_d$ via Equation~\eqref{eq:eq6},~\eqref{eq:eq7}\;
         Get Domain adapter $\boldsymbol{A}_d$ via Equation~\eqref{eq:eq1}\;
         Calculate the CTR result via Equation~\eqref{eq:eq10}\; 
   }{
   Continue\;
   }}}
   Calculate loss via Equation~\eqref{eq:eq8}\;
   Gradient calculation and update respective parameters\;
  }
\end{algorithm}
The calculation process could be tracked in Algorithm~\ref{alg:HAMUR}. Feature embeddings are obtained through the embedding layer (line 3-4), then for each instance, we select domain parameters $\boldsymbol{W}_d ^{u_l} $, $\boldsymbol{W}_d ^{u_r} $, $\boldsymbol{W}_d ^{v_l} $, $\boldsymbol{W}_d ^{v_r} $ (line 7) and domain backbone network $\mathcal{F}_d(\theta_d)$ (line 8). Afterward, we derived the adapter's parameters $\boldsymbol{U}^i_d$ and $\boldsymbol{V}^i_d$ via Equation~\eqref{eq:eq6} and Equation~\eqref{eq:eq7} (line 10-12). Then we integrate the adapter with the backbone network together to get the prediction $\hat{y}_i^d$ (line 13). Loss is calculated, followed by a parameters update (line 19-20).

\section{Experiment}

In this section, we conduct extensive experiments to verify the effectiveness of our proposed framework. Our experiments are designed to answer the following questions.
\begin{itemize}[leftmargin=*]
  \item \textbf{RQ1:} How does our proposed method compare to other state-of-the-art methods in terms of performance?
  \item \textbf{RQ2:} As the adapter is a pluggable unit, how does its performance vary when used with different backbone networks?
  \item \textbf{RQ3:} How does the size of the generated presentation matrix $I^i$ impact the performance?
  \item \textbf{RQ4:} How does the structure of the hyper-network affect the result?
\end{itemize}

\subsection{Experimental Settings}

\subsubsection{Dataset}
We conduct experiments on two public datasets. The statistics of the two datasets are shown in Table~\ref{tab:Statisticdata}.

\begin{itemize}[leftmargin=*]
  \item \textbf{MovieLens}\footnote{https://grouplens.org/datasets/movielens/}. This dataset describes people’s preferences for movies, containing 7 user features and 2 item features. We use the user feature ``age'' to divide the dataset into three different domains. Besides, we randomly split this dataset into training, validation, and test sets with a ratio of 8:1:1.
  \item \textbf{Ali-CCP} \footnote{https://tianchi.aliyun.com/dataset/408}. This dataset is collected from Alimama's traffic logs. The dataset consists of 13 user features, 5 item features, 4 combination features, and 2 label features - clicks and purchases. In this article, we only use clicks as the target label. We use the contextual feature ``301'' (representing the position) to divide the data into three domains. This dataset is already split into training, validation, and test sets by default.
\end{itemize}
\begin{table}[t]
 \caption{The statistics of datasets of each domain.}
 \label{tab:Statisticdata}
 \begin{tabular}{ccccccc}
  \toprule
  Dataset &\multicolumn{3}{c}{Ali-CCP} &\multicolumn{3}{c}{MovieLens}\\
  Domain & 1\# & 2\#& 3\# & 1\#& 2\#& 3\#\\
  \midrule
  Users & 80k&2k &136k & 1.3k & 2k & 2.6k \\
  Items &287k & 109k& 298k & 3.4k & 3.5k & 3.6k \\
  Instances &16M & 319k& 26M& 211k & 396k & 394k \\
  Persentage & 37.8\% &0.8\%&61.4\% & 21.0\% & 39.6\% & 39.4\%\\
 \bottomrule
\end{tabular}
\end{table}

\begin{table*}[ht]
 \caption{Performance Comparison Results.}
 \label{tab:ResultBaselineRUC}
 \setlength{\tabcolsep}{3mm}{
 \begin{tabular}{c|cccc|cccc}
  \toprule
  \multirow{2}{*}{Models / \textbf{AUC}} &\multicolumn{4}{c}{\textbf{MovieLens}} &\multicolumn{4}{c}{\textbf{Ali-CCP}}\\
   \cmidrule (r){2-9}
   & \multicolumn{1}{c}{1\#} & \multicolumn{1}{c}{2\#}& \multicolumn{1}{c}{3\#} & \multicolumn{1}{c|}{Total} &\multicolumn{1}{c}{1\#}&\multicolumn{1}{c}{2\#}& \multicolumn{1}{c}{3\#} & \multicolumn{1}{c}{Total}\\
  \midrule
  STAR            & 0.8052 &  0.8101   & 0.8000  & 0.8053   & 0.6302 & 0.5967 & 0.6270    & 0.6284\\
  APG            & 0.8079 &   0.8104  &0.8016   & 0.8092  & 0.6282 & 0.5962   & 0.6239      &0.6246\\
  SharedBottom   & 0.8031 &  0.8099   &0.7942   & 0.8022  & 0.6273 & 0.5908   & 0.6237     &0.6243\\
  MMOE          & 0.8074 &  0.8152  &0.8018   & 0.8086  & 0.6306 & 0.5854   & 0.6266    & 0.6280\\
  \hline
  \textbf{HAMUR}     & \textbf{0.8115*} & \textbf{0.8192*}  & \textbf{0.8030*}   &\textbf{0.8115*}  & \textbf{0.6332*} & \textbf{0.6050*}  & \textbf{0.6295*}  & \textbf{0.6300*}\\
 \bottomrule
  \toprule
  \multirow{2}{*}{Models / \textbf{LogLoss}} &\multicolumn{4}{c}{\textbf{MovieLens}} &\multicolumn{4}{c}{\textbf{Ali-CCP}}\\
   \cmidrule (r){2-9}
     & \multicolumn{1}{c}{1\#} & \multicolumn{1}{c}{2\#}& \multicolumn{1}{c}{3\#} & \multicolumn{1}{c|}{Total} &\multicolumn{1}{c}{1\#}&\multicolumn{1}{c}{2\#}& \multicolumn{1}{c}{3\#} & \multicolumn{1}{c}{Total}\\
  \midrule
  STAR         & 0.5326 &  0.5254   & 0.5308  & 0.5290   & 0.1659 & 0.2053 & 0.1595     & 0.1619\\   
  APG           & 0.5310 &   0.5269  &0.5267   & 0.5242  & 0.1678 & 0.2183   & 0.1607     &0.1630\\
  SharedBottom   & 0.5359 &  0.5258   &0.5369   & 0.5323  & 0.1658 & 0.1803   & 0.1601    &0.1624\\
  MMOE          & 0.5311 &  0.5183  &0.5291   & 0.5253  & 0.1655 & 0.1812   & 0.1601     & 0.1623\\
  \hline
  \textbf{HAMUR}         & \textbf{0.5259*} &  \textbf{0.5133*}   & \textbf{0.5245*}   & \textbf{0.5206*}  & \textbf{0.1615*} & \textbf{0.1788*}  & \textbf{0.1588*}      & \textbf{0.1614*} \\

 \bottomrule
\end{tabular}}
\\``\textbf{{\Large *}}'' indicates significance level test $p<0.05$ that compare HAMUR over the best baseline model. 
\end{table*}
\subsubsection{Baseline}
We compare our model with the multiple baselines, and they are listed as follows:
\begin{itemize}[leftmargin=*]
  \item \textbf{STAR}~\cite{sheng2021one}. STAR is a novel model with star topology that can cater to multiple domains using a unified model. It comprises shared centered parameters and multiple domain-specific parameters, which capture specific behaviors in different domains to facilitate more refined CTR prediction. The centered parameters are utilized to learn general patterns across all domains, thereby enabling the acquisition and transfer of common knowledge among all domains. 
  \item \textbf{APG}~\cite{yan2022apg}. The APG proposes an innovative approach to deep CTR models that can dynamically generate parameters for CTR prediction models, leading to more accurate results. In our comparison experiments, we choose this method as a comparison to generative parametric methods.
  \item \textbf{SharedBottom}~\cite{caruana1997multitask}. The sharedbottom model shares the bottom layer in multi-task learning. In multi-domain learning, we substitute each task tower as a domain tower.
  \item \textbf{MMOE}~\cite{ma2018modeling}. The MMOE is an approach for multi-task learning that utilizes a group of bottom networks, called  ``expert''. It is designed to gain knowledge about the fundamental feature representations. For each task, a distinct gate network is employed, and the results of the assembled experts are then passed to merge the information. For our MDR task, we treat each task tower as a domain tower and the task-specific gate network as a domain-specific gate network.
\end{itemize}

\subsubsection{Evaluation Metric}
We use two evaluation metrics Area Under the ROC \textbf{(AUC)} and \textbf{LogLoss} in our experiments to evaluate the test performance of models. Generally, a higher \textbf{AUC} or a lower \textbf{Logloss} value represents superior performance~\cite{guo2017deepfm}.

\subsection{Performance Comparison (RQ1)}

The offline comparison on MovieLens and Ali-CCP between HAMUR and baseline models are shown in Table~\ref{tab:ResultBaselineRUC}, respectively. More experiment implementation details can be found in Appendix~\ref{sec:Performance Comparison Experiment Details}. We summarize the observations as follows:

\begin{itemize}[leftmargin=*]
  \item HAMUR consistently outperforms various kinds of state-of-the-art models over all datasets by a significant margin, demonstrating the effectiveness of HAMUR in MDR tasks. Compared with the baseline model STAR, HAMUR utilizes a hyper-network to share domain information rather than directly multiplying a shared fully connected network. HAMUR is capable of capturing domain characteristics with flexibility. In contrast to APG, which generates weights for the backbone network, HAMUR generates parameters for each domain-specific adapter, thereby retaining the specialties of each domain. Compared with the multi-task models SharedBottom and MMOE, HAMUR explicitly models the domain information better to capture the distinct distribution characteristics between different domains.
  \item HAMUR can boost the performance of models on different domains, especially in the domain where training samples are rare, such as the 2\# domain in Ali-CCP. As our model operates at an instance level, it can effectively incorporate enough information from other domains, ultimately improving overall performance.
  \item Our model has demonstrated promising performance across two different datasets, which can be attributed to the efficiency of the hyper-network. The hyper-network is adept at capturing fine-grained domain similarity at the instance level, thereby enabling effective characterization of data distribution characteristics even for disparate datasets.
\end{itemize}

\subsection{Compatibility Experiment (RQ2)}

As mentioned above, our proposed HAMUR is a pluggable component to different backbone networks. 
So, in this section, we choose three different networks MLP~\cite{amari1967theory}, DCN~\cite{wang2017deep}, Wide \& Deep~\cite{cheng2016wide}, and the result is shown in Table~\ref{tab:Different Backbones}. The experiment comprises an independent backbone network test and a backbone network with our proposed HAMUR test, akin to the ablation experiment setup. The performance of the model experiment is reported on two datasets. For more experiment implementation details, please refer to Appendix~\ref{sec:Compatibility Experiment Details}. We summarize the observations as follows:

\begin{table*}[ht]
 \caption{Compatibility Experiment Results.}
 \label{tab:Different Backbones}
 \setlength{\tabcolsep}{4mm}{
 \begin{tabular}{c|ccc|ccc}
  \toprule
  \multirow{2}{*}{Models / \textbf{AUC}} &\multicolumn{3}{c}{\textbf{MovieLens}} &\multicolumn{3}{c}{\textbf{Ali-CCP}}\\
   \cmidrule (r){2-7}
   & \multicolumn{1}{c}{1\#} & \multicolumn{1}{c}{2\#}& \multicolumn{1}{c|}{3\#} &\multicolumn{1}{c}{1\#}&\multicolumn{1}{c}{2\#}& \multicolumn{1}{c}{3\#} \\
  \midrule
  MLP            & 0.8105 &  0.8171   &0.8021   & 0.6312  & 0.5821   & 0.6250    \\
  MLP+\textbf{HAMUR}   & \textbf{0.8115} &   \textbf{0.8192}  &\textbf{0.8030}   & \textbf{0.6332}  & \textbf{0.6050}  & \textbf{0.6295}   \\
  \hline
  DCN           &  0.8141   & 0.8184  & 0.8025   & 0.6305 & 0.6059 & 0.6270    \\
  DCN+\textbf{HAMUR}   & \textbf{0.8165} &  \textbf{0.8201}  & \textbf{0.8061}   & \textbf{0.6319}  & \textbf{0.6064} & \textbf{0.6297}   \\
  \hline
  Wide \& Deep                  & 0.8099 &  0.8137   & 0.7997  & 0.6254   & 0.5963 & 0.6221 \\
  Wide \& Deep+\textbf{HAMUR}          & \textbf{0.8131} &  \textbf{0.8172}  & \textbf{0.8035}   & \textbf{0.6306}  & \textbf{0.6053} & \textbf{0.6262} \\
 \bottomrule
   \toprule
  \multirow{2}{*}{Models / \textbf{LogLoss}} &\multicolumn{3}{c}{\textbf{MovieLens}} &\multicolumn{3}{c}{\textbf{Ali-CCP}}\\
   \cmidrule (r){2-7}
   & \multicolumn{1}{c}{1\#} & \multicolumn{1}{c}{2\#}& \multicolumn{1}{c|}{3\#} &\multicolumn{1}{c}{1\#}&\multicolumn{1}{c}{2\#}& \multicolumn{1}{c}{3\#} \\
  \midrule
  MLP            & 0.5270 &  0.5162   &0.5298   & 0.1625  & 0.2124   & 0.1599   \\
  MLP+\textbf{HAMUR}   & \textbf{0.5259} &   \textbf{0.5133}  & \textbf{0.5245}  & \textbf{0.1615}  & \textbf{0.1788}  & \textbf{0.1588}   \\
  \hline
  DCN           &  0.5231   & 0.5136  & 0.5238   & 0.1659 & 0.1802 & 0.1601     \\
  DCN+\textbf{HAMUR}   & \textbf{0.5180} &  \textbf{0.5107}  & \textbf{0.5211}   & \textbf{0.1657}  & \textbf{0.1800} & \textbf{0.1596}   \\
  \hline
  Wide \& Deep              & 0.5313 &  0.5215   & 0.5319  & 0.1648   & 0.1796 & 0.1594 \\
  Wide \& Deep+\textbf{HAMUR}     & \textbf{0.5258} &  \textbf{0.5173}  & \textbf{0.5273}   & \textbf{0.1631}  & \textbf{0.1788} & \textbf{0.1588} \\
 \bottomrule
\end{tabular}}
\end{table*}
\begin{itemize}[leftmargin=*]
  \item Our proposed HAMUR is notably capable of comprehending domain distinctions and effectively incorporating domain representations into existing backbone networks, leading to improved prediction results across all different backbone models. This advantage could be attributed to the design of the adapter. The domain-specific adapter can significantly change the distribution of the original domain.
  \item Besides, our results demonstrate that HAMUR can consistently improve the performance of different types of backbone models across various datasets, underscoring its efficacy and potency. We attribute this to our proposed hyper-network. Because it implicitly captures domain features for different distributions.
  \item Moreover, for different backbones, The lift that our model brings varies. For example, wide \& deep improve significantly on Ali-CCP after combining our model. We attribute this as the consequence of the inner bias of different models. Given that our network achieves excellent migration performance, other backbones that would like to expand into multiple domain models do not necessitate a complete restructure. Integrating our HAMUR model alone can produce outstanding performance. Our approach holds promising implications for practical implementation.
\end{itemize}

\subsection{Hyper Parameters Experiment (RQ3 \& RQ4)}

\subsubsection{Dimension of Presentation Matrix (RQ3)}

\begin{figure}[ht]
  \centering
 \includegraphics[width=0.485\linewidth]{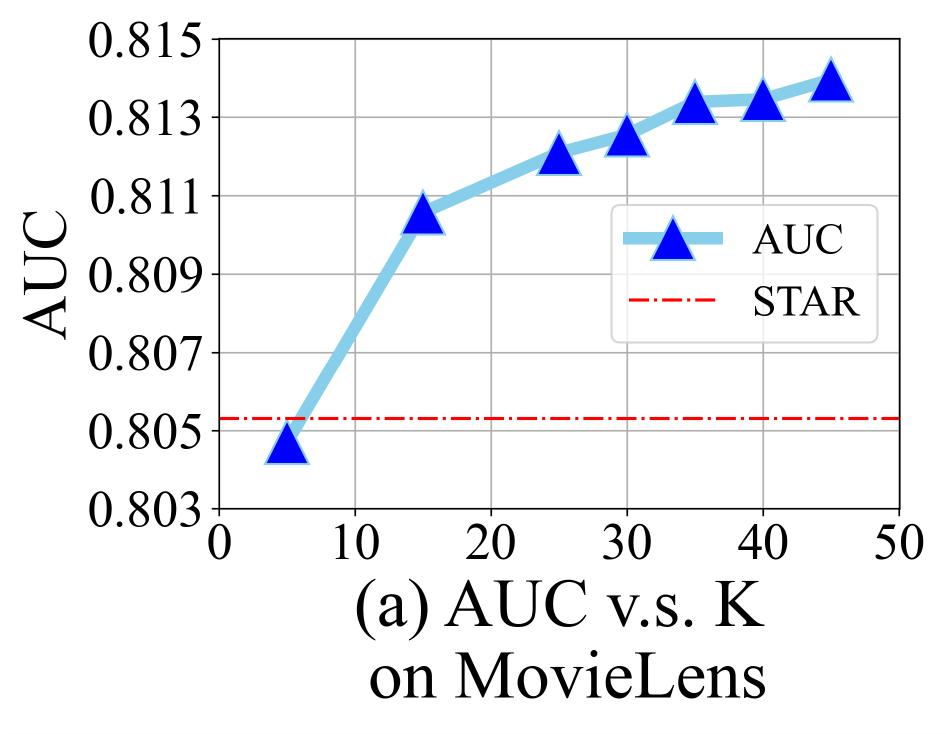}
 \includegraphics[width=0.485\linewidth]{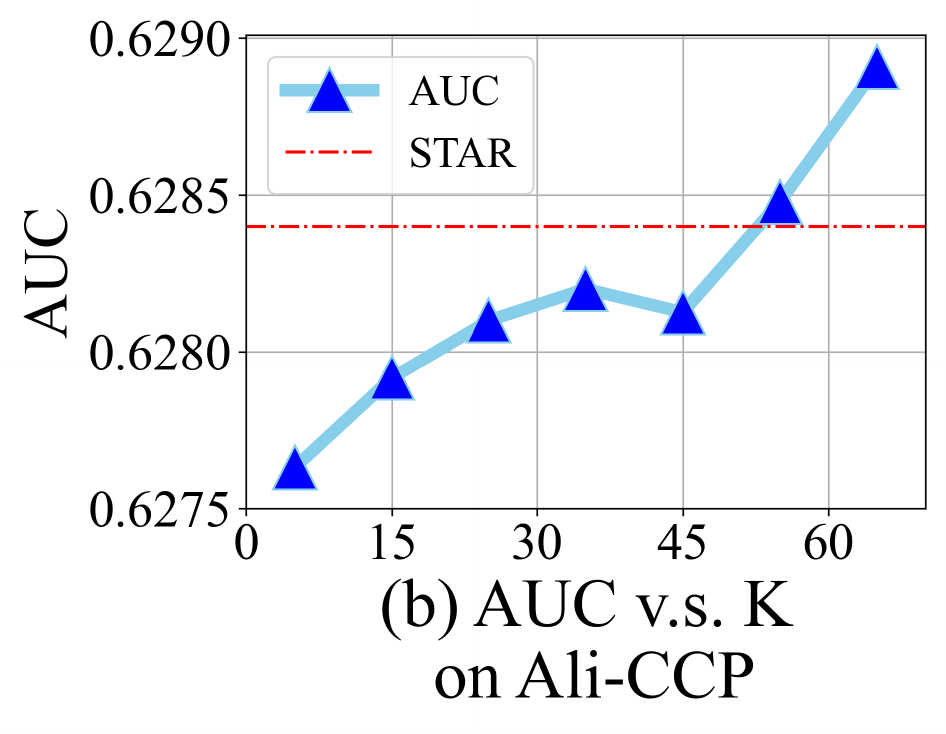}

 \caption{The effect of $K$ on AUC on different datasets.}
 \Description{ AUC with $K$ on Ali-CCP}
 \label{fig:k}
\end{figure}

As discussed in Section~\ref{sec:Domain Shared Hyper Network}, utilizing low-rank dimension allows for decomposing the target matrices $\boldsymbol{U}$ and $\boldsymbol{V}$ into three matrix multiplications. We only need to generate the representation matrix $\boldsymbol{I}^i \in \mathbb{R}^{k \times k}$ for each instance. The value of $k$ determines the size of matrix $\boldsymbol{I}^i$ thus impacts the representation ability of the hyper-network. To investigate the effect of $k$ on the performance of our model, we conducted experiments with varying values of $k$. The results are presented in Figure~\ref{fig:k}.

Concretely, we conducted a performance comparison experiment of various $k$ values against the baseline model STAR. From Figure~\ref{fig:k}~(a) and Figure~\ref{fig:k}~(b), we can see that even for different datasets, the curve shows the same trend, revealing that as $k$ increases, the model's effectiveness improves. STAR exhibited a slight advantage over HAMUR at lower $k$ values, but our model achieved significant AUC enhancements as $k$ values increased. A larger $k$ is needed to exceed STAR in Ali-CCP. We assume this is caused by the extremely uneven distribution of Ali-CCP in each domain. However, given that our model operates at the instance level and generates a matrix for each instance, choosing an appropriate $k$ value is crucial to avoid the excessive computational burden. Therefore, selecting a suitable value of $k$ within the limits of available computing resources is necessary.

\subsubsection{Dimension of Hyper-network (RQ4)} 
\begin{figure}[h]
  \centering
 \includegraphics[width=0.485\linewidth]{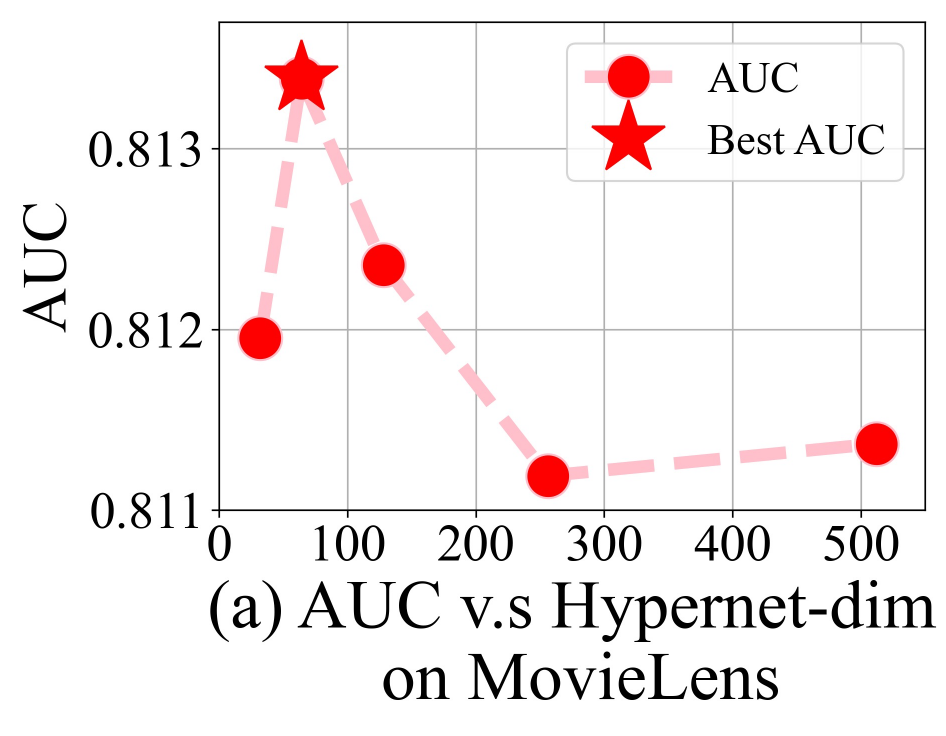}
 \includegraphics[width=0.485\linewidth]{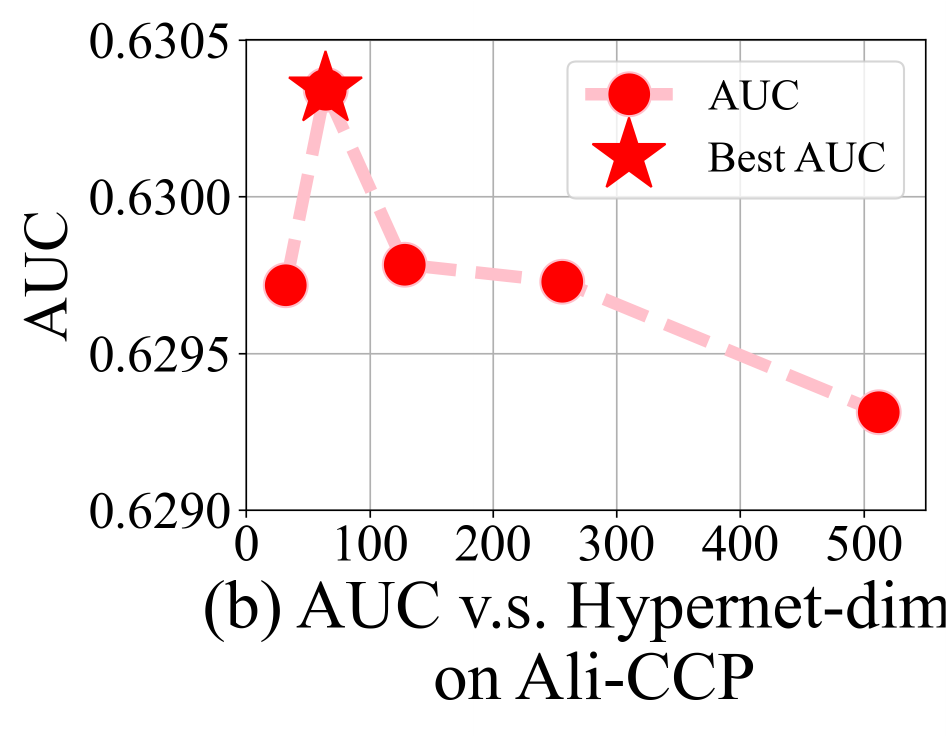}
 
 \caption{{The effect of hyper-network dimensions on different datasets.}
 }
 \label{fig:dims}
\end{figure}

As discussed in Section~\ref{subsec:Parameters Generation}, the architecture of the hyper-network is a shallow neural network. To be exact, it constitutes a single-layered Deep Neural Network. To investigate the impact of various dimensions of the hidden state of the hypernet on the model's performance, we conducted experiments on two datasets, and the outcomes are presented in Figure~\ref{fig:dims}.

The findings in two public datasets reveal that augmenting the dimensionality of the hyper-network does not improve the model's efficacy. Actually, there is a decline in performance after reaching the peak, demonstrating that hyper-network is susceptible to overfitting. Mere augmentation of dimensionality fails to enhance the expressive potential of the hyper-network, instead, yields detrimental outcomes. This hints that there is no need to choose a significant dimension of the hyper-network.

\section{Related Work}
This section will briefly introduce some topics related to our study. We will first introduce the recent studies of MDR and followed by a brief review of dynamic weight generation in deep neural networks. The latest research and relevant literature on adapters will be presented at the end of this section.

\subsection{Multi-Domain Recommendation}
The MDR model is an approach that seeks to enhance the performance of recommendation systems models across multiple domains by constructing a unified model that serves all domains simultaneously. In contrast to the cross-domain recommendation approaches ~\cite{li2020ddtcdr, hu2018conet,gao2023autotransfer,wang2023single,fan2023adversarial}, cross-domain recommendation models only focus on transferring knowledge from a single source domain to a target domain, however, MDR models seek to optimize performance across all domains. 

Various methods have been proposed to address multi-domain-related problems in recommendation systems. We categorize the present strategies into different perspectives. (1). From the perspective of causal inference. CausalInt~\cite{wang2022causalint} proposed TransNet to simulate a counterfactual intervention and integrate domain information. (2). From the perspective of embedding. AdaptDHM~\cite{li2022adaptdhm} proposed a distribution adaptation module that captures the commonalities and distinctions between distinct embedding clusters, thus executing different domain strategies. (3). From the perspective of pruning. Adasparse~\cite{yang2022adasparse} proposed a domain pruner, which takes different pruning methods for different domains. (4). From the Generative Adversarial Network (GAN) perspective. AFT~\cite{hao2021adversarial} proposed a domain-specific masked encoder to capture domain-related feature interactions in the generator. In the multi-domain discriminator, AFT utilizes knowledge representation learning to model relations between items, users, and domain indicators. (5). From the perspective of the attention mechanism. SAR-Net~\cite{shen2021sar} proposes attention layers to capture users’ cross-domain interest transfer and design a domain-specific and domain-shared experts network pipeline. (6). From the perspective of topology. STAR~\cite{sheng2021one} proposed a star topology with a shared fully connected network at the center, which is multiplied by domain-specific fully connected networks in the domain-specific calculation. (7). From the perspective of meta-learning. M2M~\cite{zhang2022leaving} proposed a meta-unit to simultaneously learn domain-related and task-related information to achieve multi-domain multi-task learning. 

Unlike the above methods, our proposed method is a pluggable component with strong scalability that could be combined with multiple backbone models.

\subsection{Dynamic Weight Generation In Deep Neural Networks}

Weight generation originated from evolutionary computing, which addresses the challenge of operating in large search spaces comprising millions of weight parameters. It is challenging to manipulate such vast search spaces directly. Therefore, weight generation was developed as an alternative approach to overcome this limitation. Hyper-network~\cite{ha2016hypernetworks} was proposed using a small network to generate weights for RNN, achieving outstanding performance. These weight-generation methods have been used in Other areas like computer vision~\cite{ha2019end, mok2021conditional}, nature language processing~\cite{dangovski2019rotational, mahabadi2021parameter}, reinforcement learning~\cite{rezaei2023hypernetworks, beck2023hypernetworks}, speech~\cite{du2021meta}.

Dynamic weight generation approaches have also been widely used in the context of recommendation systems. CAN~\cite{bian2020can} proposes a co-action unit in which the items model's weights are generated dynamically. APG~\cite{yan2022apg} generates different models' weight matrices and gets great performance. M2M~\cite{zhang2022leaving} generates meta-units' weight matrix with scenario knowledge. Compared with these approaches, we follow the same paradigm of parameter generation, but in our method, we generate parameters for the adapters. 
\subsection{Adapter}
The adapter~\cite{rebuffi2017learning} module was originally proposed in the field of computer vision, which is developed as a tunable deep neural network architecture that can be steered on the fly to diverse visual objects. This module was deployed and used in nature language processing by~\citep{houlsby2019parameter}. Transfer learning was initially employed through pre-trained feature vectors and fine-tuning of pre-trained language models (PLMs). The adapter presents a novel approach: integrating a small number of parameters into the model, which can be trained only during the fine-tuning of downstream tasks while keeping the parameters of the pre-training model unaltered. 

The existing adapters can be divided into the following three types, series adapters, parallel adapters, and Low-Rank Adaption models (LoRA). In the series adapter, the adapter cell is added in series behind each multi-head attention layer and a feed-forward layer of the transformer block. The parallel adapter connects adapter cells in parallel to each multi-head attention layer and feed-forward layer. Whereas LoRA introduces trainable low-rank factorization matrices in the existing layers of LLMs, enabling the model to adapt to new data while keeping the original LLMs fixed to preserve existing knowledge. This approach offers several benefits. (1). The requirement for training only a small number of parameters of the adapter module leads to reduced training costs and enhanced portability. (2). For continual learning of different tasks, adapters solely need to be trained, while other parameters remain unaltered, thereby preventing forgetting past knowledge when learning new tasks.

The adapter has demonstrated remarkable performance across diverse tasks. Its utilization in neural machine translation~\cite{bapna2019simple}, relation and classification~\cite{wang2020k}, entity typing~\cite{wang2020k}, and even multimodal tasks~\cite{lin2023multimodality} have yielded impressive outcomes. In our paper, we proposed HAMUR. As far as we know, it is the first time that the adapter has been used in MDR and adapts to the end-to-end training method of the recommendation system.
\section{Conclusion}

In this paper, we proposed a novel model, Hyper Adapter Multi-Domain Recommendation Model (HAMUR) for Multi-Domain Recommendation (MDR). HAMUR comprises two components, namely, a domain-shared hyper-network that dynamically captures domain-shared information and a domain-specific adapter that can be easily integrated into different existing networks. The model offers several innovative features: firstly, it utilizes a hyper-network to implicitly capture domain-shared information, thereby avoiding the domain-bias problem associated with hard-share-parameters methods. Secondly, to the best of our knowledge, HAMUR is the first model to use an adapter in MDR, and the adapter is directly integrated into the model to achieve end-to-end training. Lastly, extensive experiments on two public datasets demonstrate the superior performance of HAMUR. Future research will investigate the model in fine-grained domains and verify its feasibility in more complex multi-task scenarios.

\begin{acks}
This research was partially supported by Huawei (Huawei Innovation Research Program), APRC - CityU New Research Initiatives (No.9610565, Start-up Grant for New Faculty of City University of Hong Kong), CityU - HKIDS Early Career Research Grant (No.9360163), Hong Kong ITC Innovation and Technology Fund Midstream Research Programme for Universities Project~(No. ITS/034/22MS), SIRG - CityU Strategic Interdisciplinary Research Grant (No.7020046, No.7020074), SRG-Fd - CityU Strategic Research Grant (No.7005894). We thank MindSpore~\cite{mindspore} for the partial support of this work, which is a new deep learning computing framework.
\end{acks}
\appendix
\section{Experiment Details}
\subsection{Performance Comparison Experiment Details \label{sec:Performance Comparison Experiment Details}}
The following setups are adopted to ensure a fair comparison in the baseline experiment.
\begin{itemize}[leftmargin=*]
  \item Backbone models of STAR, APG, and HAMUR are implemented using the same dimension of MLP. Specifically, for the MovieLens dataset, a 2-layer MLP is chosen. For the Ali-CCP dataset, a 7-layer MLP is selected.
  \item The dimension of the hyper-network in HAMUR, the auxiliary network in STAR, and the expert network are set to be identical without loss of generality.
  \item The location of the adapter that has been plugged is situated on the top layer of the base model. Conctreatly, in the dataset of MovieLens, the position is situated between the first and second layers. But in the dataset of Ali-CCP, we insert two adapter cells at the top layer of the base model.
\end{itemize}
For details, refer to Table~\ref{tab:baselineParameters}.
\begin{table}[t]
 \caption{The Hypter parameters of Baseline Comparison Experiment.}
 \label{tab:baselineParameters}
 \begin{tabular}{ccc}
  \toprule
  Parameters & Movielens & Ali-CCP\\
  \midrule
  Embedding Size & 16&16 \\
  Bottole Neck Dimension ($h$) &32 &32 \\
  Hyper-Net Dimension & 64 & 64\\
  Representation Matrix Dimension ($k$) & 35 & 65\\
  Backbone Net Layers & 2 & 7\\
  
\bottomrule
\end{tabular}
\end{table}

\begin{figure}[t]
  \centering
 \includegraphics[width=\linewidth]{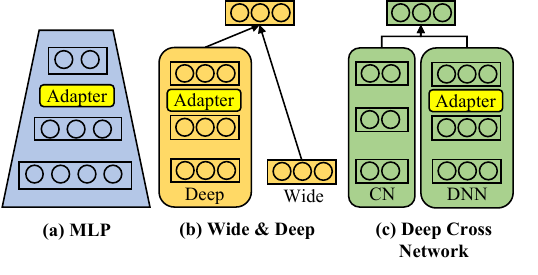}
 \caption{Different Backbone Network.}
 \Description{An illustration of the different backbone with HAMUR.}
 \label{fig:differentbackbone}
\end{figure}

\subsection{Compatibility Experiment Details\label{sec:Compatibility Experiment Details}}

For Different backbone network experiments, we compare three networks: MLP, DCN, and Wide \& Deep. The general structures are shown in Figure~\ref{fig:differentbackbone}. The following setups are employed for conducting the experiments.
\begin{table}[t]
 \caption{The Hypter parameters of Compatibility Experiment.}
 \label{tab:backboneParameters}
 \setlength{\tabcolsep}{0.15mm}{
 \begin{tabular}{cccc}
  \toprule
  Model & Parameters & Movielens & Ali-CCP\\
  \midrule
  \multirow{5}{*}{DCN}
  &Embedding Size & 16&16 \\
  &Bottole Neck Dimension ($s$) &32 &32 \\
  & Number of Cross Layers & 2&7 \\
  &Hyper-Net Dimension & 128 & 64\\
  &Representation Matrix Dimension ($k$) & 30 & 25 \\
  &Deep Network Layers & 2 & 6 \\
  
  \midrule
  
  \multirow{5}{*}{Wide \& Deep}
  &Embedding Size & 16&16 \\
  &Bottole Neck Dimension ($s$) &32 &32 \\
  &Hyper-Net Dimension & 128 & 64\\
  &Representation Matrix Dimension ($k$) & 45 & 25 \\
  &Deep Side Layers & 2 & 6 \\
\bottomrule
\end{tabular}}
\end{table}
\begin{itemize}[leftmargin=*]
  \item We design two kinds of experiments, integrating HAMUR in the backbone network and without HAMUR in the backbone network. Taking DCN as an example, each domain will construct a DCN independently as a backbone network. In the experiment ingratiating the backbone network and HAMUR, domain-specific adapters will be inserted in each backbone network, and a domain-shared hyper-network will also be constructed to capture domain-shared information.
  \item The adapter is integrated into the Deep Network for DCN and into the deep side for Wide \& Deep, and for MLP, we follow the setting elaborated in the Appendix~\ref{sec:Performance Comparison Experiment Details}.
  \item Various insertion positions of the adapter are tested through extensive experiments, and it is observed that the positions closer to the model's terminus produced the most favorable performance. Consequently, all adapters are integrated between the last and penultimate layers.
\end{itemize}
For details, refer to Table~\ref{tab:backboneParameters}.

\bibliographystyle{ACM-Reference-Format}
\balance
\bibliography{7Reference}

\end{document}